\documentclass{epl}
\newcommand {\bb}{\bibitem}
\newcommand {\be}{\begin{equation}}
\newcommand {\ee}{\end{equation}}
\newcommand {\bea}{\begin{eqnarray}}
\newcommand {\eea}{\end{eqnarray}}

\title{Upper critical field H$_{c2}$ in PrOs$_{4}$Sb$_{12}$}
\author{D. Parker\inst{1,2} \and K. Maki\inst{1,2} \and
H. Won\inst{1,3}}
\institute{\inst{1} Max-Planck Institute for the Physics of Complex Systems,
N\"othnitzer Str. 38 D-01187, Dresden, Germany\\
\inst{2} Department of Physics and Astronomy, University of Southern
California, Los Angeles, CA 90089-0484 USA\\
\inst{3} Department of Physics, Hallym University,
Chuncheon 200-702, South Korea}

\begin{document}
\maketitle
\begin{abstract}
We study the upper critical field of the A and B phases in the
triplet superconductor PrOs$_{4}$Sb$_{12}$
within the p+h-wave superconductivity proposed recently for this material.
The present result is compared with H$_{c2}$(t) and H*(t), the boundary
between the A and B phase in PrOs$_{4}$Sb$_{12}$, reported earlier and with 
more recent data of H$_{c2}(t)$ for the single phase crystal.  We find
H$_{c2}$(t)'s for both the two phase crystal and the single phase crystal are 
described by the model for the A phase.  From this fitting one can deduce
the Fermi velocity as $v=2.5 \times 10^6$ cm/s.  On the other hand H$_{c2}$(t)
for the B phase is found to be somewhat smaller than H*(t), which is rather
puzzling.
\end{abstract}

\section{Introduction}

Superconductivity in the filled skutterudite PrOs$_{4}$Sb$_{12}$ discovered
in 2002 by Bauer et al \cite{1,2,3} has generated a big sensation.  First
the presence of at least two distinct phases (A and B phases) in 
a magnetic field, with both gap functions having point nodes is surprising.  
Further these 
superconductors belong to the triplet pairing with broken chiral symmetry
\cite{4,5,6}.  However, the exact location of the A-B phase boundary is 
still controversial \cite{7}.  More recently Measson et al 
\cite{measson,measson2} have discovered many PrOs$_{4}$Sb$_{12}$ crystals
with only a single phase.  Therefore the nature of this single phase has to be 
addressed.  In this paper we assume that the gap functions
of the A and B phases of PrOs$_{4}$Sb$_{12}$ are given as \cite{6} 
\bea
{\bf \Delta}_{A}({\bf k}) &=& {\bf d_{A}} e^{\pm i\phi_{i}}\frac{3}{2}(1-\hat{k}_{x}^{4}-\hat{k}_{y}^{4}-\hat{k}_{z}^{4})).\\
{\bf \Delta}_{B}({\bf k}) &=& {\bf d_{B}} e^{\pm i\phi_{3}}(1-\hat{k}_{z}^{4})
\eea
Here $e^{\pm i\phi_{i}}$ is one of $e^{i\phi_{1}}=( \hat{k}_{y}+i \hat{k}_{z})/\sqrt{\hat{k}_{y}^{2}+\hat{k}_{z}^{2}},
e^{i\phi_{2}}=(\hat{k}_{z}+i\hat{k}_{x})/\sqrt{\hat{k}_{z}^{2}+\hat{k}_{x}^{2}},
e^{i\phi_{3}}= (\hat{k}_{x}+i\hat{k}_{y})/\sqrt{\hat{k}_{x}^{2}+\hat{k}_{y}^{2}}$. 
The factor of 3/2 ensures proper normalization of the angular dependence of the order parameter.
In Eq.(2) the nodal direction is chosen to be parallel to (001).  We note that
$|\Delta_{A}({\bf k})|$ retains 
the cubic symmetry while $|\Delta_{B}({\bf k})|$ 
has only the axial symmetry.  In the absence of an external perturbation, we
assume that the nodal direction of $|\Delta_{B}({\bf k})|$ is parallel to
${\bf H}$, the magnetic field, when ${\bf H}$ is parallel to one of the 
crystal axes.  Although many other gap functions for PrOs$_{4}$Sb$_{12}$ have
been proposed so far, none can describe the thermal conductivity data
of Izawa et al as discussed in \cite{4,6}.  For a discussion of our
methodology for determining the gap functions of nodal superconductors
please see Ref. \cite{sal}.

In the following we study the upper critical field H$_{c2}$(T) of the A
and B phases of PrOs$_{4}$Sb$_{12}$ 
in terms of the linearized gap equation \cite{gorkov}.
For unconventional superconductors Gor'kov's original formulation must be
modified as shown by Luk'yanchuck and Mineev \cite{8}.  The upper critical
field thus obtained for the A phase describes consistently H$_{c2}$ of the
A phase as obtained by Measson et al \cite{7}.  On the other hand H*(T) is
somewhat larger than H$_{c2}$(T) in the B phase at low temperatures.

Of course H*(T) is not exactly equal to H$_{c2}$(T) of the B phase.  
But it is rather puzzling that H*(T) is larger than the predicted 
H$_{c2}$(T) of the B phase.  Meanwhile, some of the single crystals
exhibit only a single phase transition.  Measson has reported H$_{c2}$(T) of
the single phase system at SCES05 in Vienna \cite{measson}.  The observed
H$_{c2}$(T) is practically the same as H*(T), and consistent with
the one for the A phase.  This, perhaps, implies that the A phase is more
robust than the B phase.  Also, perhaps H*(T) is associated with the A phase.

From the spatial configuration of $\Delta({\bf r,k})$ at H=H$_{c2}$(T), we can
deduce the stable vortex lattice structure as in \cite{9,10}.
Except in the immediate
vicinity of the superconducting transitions ($T < 0.95 T_{c}$) we find 
the square vortex lattice is more stable than the hexagonal one
for ${\bf H} \parallel$ [001] or any of the cubic axes, while the 
hexagonal lattice is more stable for ${\bf H} \parallel$ [111] for example.
Recently the vortex lattice in PrOs$_{4}$Sb$_{12}$ was studied via small
angle neutron scattering (SANS) by Huxley et al \cite{16}.  They find only
a distorted hexagonal lattice at T=100 mK and H =0.1 T, most likely due to
poor sample quality.

\section{Upper critical field}

As in the chiral p-wave superconductor the upper critical field is
determined from the linearized gap equation for ${\bf \Delta(r,k)}$:
\bea
{\bf \Delta(r,k)} \sim (e^{i\phi}+ Ce^{-i\phi}(a^\dagger)^2)|0>
\eea
where $|0>$ is the Abrikosov state \cite{11}:
\bea
|0> &=& \sum_{n} c_{n}e^{-eBx^{2}-nk(x+iy) - \frac{(kn)^{2}}{4eB}}
\eea
and $a^{\dagger} = \frac{1}{\sqrt{2eB}}(-i\partial_{x}-\partial_{y}+
2ieBx)$ is the raising operator and we have assumed ${\bf H} \parallel 
{\bf c}$.

Then following Refs. \cite{9,10} the upper critical field is determined by
\bea
-\ln t = \int_{0}^{\infty} \frac{du}{\sinh u}(1-<f^{2}>^{-1}\langle f^{2}
e^{-\rho u^{2}(1-z^{2})} \times (1 +2C\rho u^{2}(1-z^{2}))\rangle) \\
-C \ln t = \int_{0}^{\infty} \frac{du}{\sinh u}(1-<f^{2}>^{-1}\langle f^{2}
e^{-\rho u^{2}(1-z^{2})} \times (\rho u^{2}(1-z^{2}) + \nonumber \\
C(1 - 4\rho u^{2}(1-z^{2}) +2\rho^{2}u^{4}(1-z^{2})^{2})\rangle))
\eea
where $t=\frac{T}{T_{c}}, \rho=\frac{v^{2}eH_{c2}(t)}{2(2\pi T)^{2}}$, where
v is the Fermi velocity and 
\bea
f &=& 1- \cos^{4}\theta - \sin^{4}\theta(\sin^{4}\phi+\cos^{4}\phi) 
\mathrm{,\,\,\,(A-phase)} \\
&=& 1- \cos^{4}\theta \mathrm{,\,\,\,(B-phase)} 
\eea
and $\langle \ldots \rangle$ means
$\int d\Omega/(4\pi) \ldots$.  Also $<f^{2}> 
= \frac{4}{21}$ and $\frac{32}{45}$ 
for the A and B phases respectively.  Then for the A phase we obtain the
following asymptotics for $t \simeq 1$:
\bea
-\ln t &=& \frac{3\zeta(3)}{7}\rho(1-2C) \mathrm{\,\,\,and} \\
-C\ln t &=& \frac{3\zeta(3)}{7}\rho(-1+5C)
\eea
where $\zeta(3)=1.202...$.  From this we find $C=\sqrt{\frac{3}{2}}-1$ and
$\rho=\frac{3}{7\zeta(3)}(1+\sqrt{\frac{2}{3}})(-\ln t) = 0.5942 (-\ln t)$.
On the other hand, for t=0 we find
\bea
0 &=& \ln(4\gamma \rho_{0})+ <f^{2}>^{-1}\langle f^{2}\ln(1-z^{2})\rangle - 2C 
\mathrm{\,\,\,and} \\
0 &=& C(\ln(4\gamma \rho_{0})+ <f^{2}>^{-1}\langle f^{2}\ln(1-z^{2})+2\rangle - 1
\eea
where $\gamma=1.780...$ is the Euler constant and $<f^{2}>^{-1}<f^{2}\ln(1-z^{2})> = -0.40104$.  These yield $C=\frac{1}{2}(\sqrt{3}-1)$ and
\bea
\rho_{0} & \equiv & \frac{v^{2}eH_{c2}(0)}{2(2\pi T_{c})^{2}} =\frac{1}{4\gamma}\exp[1.2295] = 0.4795
\eea
From this we obtain
\bea
h(0) &=& \frac{H_{c2}(0)}{-\frac{dH_{c2}(t)}{dt} |_{t=1}} =
0.80696
\eea
H$_{c2}(t)$ and $C(t)$ are evaluated numerically and shown in Fig. 1 and Fig. 2
respectively.  Also in Fig. 1 
H$_{c2}$(t) from the A phase is compared with the data 
from Ref. \cite{7}.
\begin{figure}[h]
\includegraphics[width=8cm]{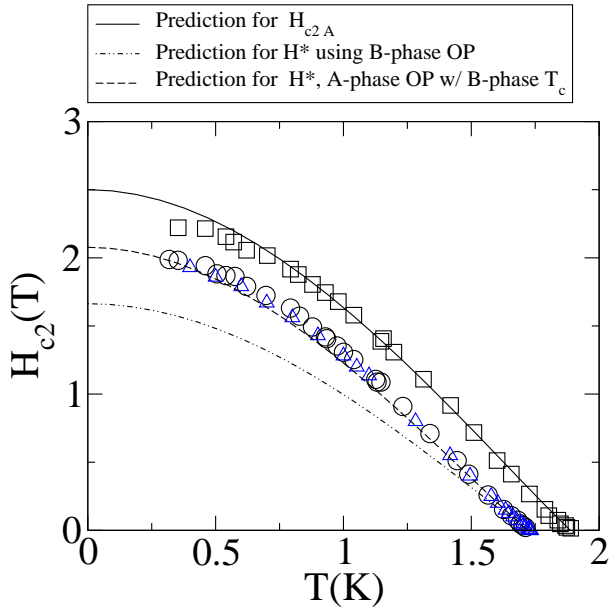}
\caption{H$_{c2}$(T) for the A and B phases is compared with the experimental
data for H$_{c2}$(T) and H*(T).  The triangles indicate the single phase data.}
\end{figure}
As is readily seen, we have an excellent
fit, where we used $v = 2.45 \times 10^{6}$ cm/sec
which is very reasonable.  This v is somewhat smaller than
the one deduced from the thermal conductivity data \cite{6}.
However, it is known that the thermal conductivity data in the clean limit
(and not in the superclean limit) is less sensitive to the actual
value of v.  In the same way v is smaller than that deduced from de Haas-van
Alphen data \cite{14}.  But in heavy-fermion systems such as UPt$_{3}$, it
is well known that dHvA gives a lighter effective mass than that deduced
via the specific heat.

For the B phase we find for $t \simeq 1$
\bea
-\ln t &=& \frac{31\zeta(3)}{11}\rho(1-2C) \mathrm{\,and} \\
-C\ln t &=& \frac{31\zeta(3)}{11}\rho(-1+5C)
\eea
which give $C= \sqrt{\frac{3}{2}}-1$ and
$\rho=\frac{11}{31\zeta(3)}(1+\sqrt{\frac{2}{3}})(-\ln t) \simeq 0.5362 (-\ln t)$.
For $ t \simeq 0$, on the other hand, we have the same set of equations as in
Eqs.(10) and (11), where
\bea
<f^{2}>^{-1}<f^{2}\ln(1-z^{2})> = -0.25973
\eea
This gives $C= \frac{1}{2}(\sqrt{3}-1)$ and
\bea
\rho_{0} &=& \frac{1}{4\gamma}\exp[1.2295025] = 0.3786
\eea
We then find
\bea
h(0) &=& \frac{H_{c2}(0)}{-\frac{d{H_{c2}(t)}}{dt}|_{t=1}} =
0.7062115
\eea
Both H$_{c2}(t)$ and C(t) for the B phase are obtained numerically and
shown in Fig. 1 and Fig. 2 respectively as before.  Also H*(t) from
\begin{figure}[h]
\includegraphics[width=8cm]{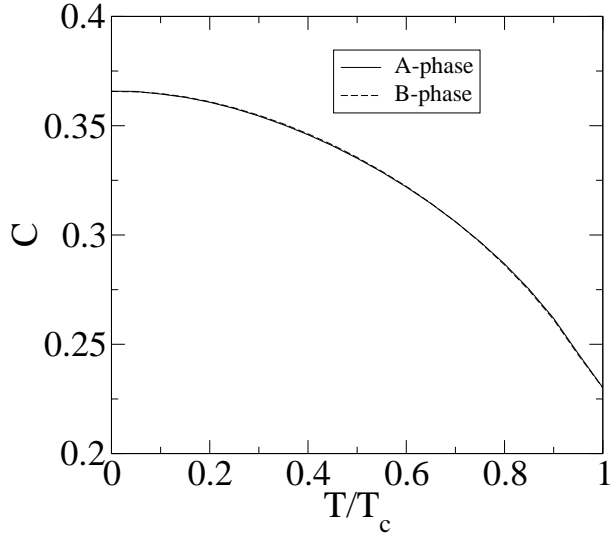}
\caption{C(T) for the A and B phases.}
\end{figure}
Ref. [7] is shown in Fig. 1 as well, though there is
no reason that H*(t) should correspond to H$_{c2}$(t) of the B phase.
Indeed we see that H$_{c2}$(t) for the B phase is substantially smaller
than the observed H*(t), especially at low temperatures.

We also compare in Fig. 1 the observed H$_{c2}$(t) of the single phase crystal 
with previous data, as well as our predictions for the A and B phases.  The
single phase data appears essentially identical to H*(t) of the two phase data,
suggesting that the single-phase crystals have only the A phase.  
H*(t) and H$_{c2}$(t) of the single phase crystal are fully
consistent with H$_{c2}$(t) of the A phase, if T$_{c}$ is taken as 1.72 K,
as indicated by the middle fit line in Fig. 1.  Such a reduction in T$_{c}$
is readily reproduced in the presence of impurity scattering \cite{8}.  Also we recall
the scaling of H$_{c2}(T/T_{c},\Gamma)$/H$_{c2}(0,\Gamma)$ in the presence of 
impurity scattering.  
These strongly suggest that both H*(t) and H$_{c2}$(t) of the
single phase system should be associated with the A phase.  As we have
mentioned earlier the actual A-B phase boundary is still controversial; below 
in Fig. 3 we show a measurement \cite{4,6} indicating a phase
\begin{figure}[h!]
\includegraphics[width=8cm]{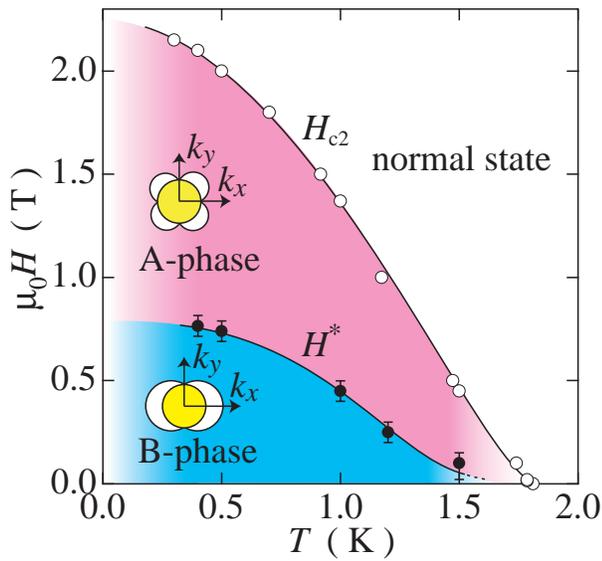}
\caption{Alternate phase diagram of PrOs$_{4}$Sb$_{12}$}
\end{figure}
boundary lying much lower in the H-T plane, 
and hence more consistent with our preduction
for H$_{c2}(t)$ of the B phase.  Clearly experiments with cleaner
crystals are desirable.

Also from Fig. 2 we notice that C(t) for the B phase is practically the
same as the one obtained for the A phase.  Further the present C(t) is
practically the same as C(t) obtained for 
the chiral p-wave superconductor \cite{9}.  We believe there should be an
analytical reason that C(t) is universal for the class of chiral
superconductors.  This means that the Abrikosov state consists of the n=0
Landau wave function with an admixture of the n=2 Landau wave functions
with opposite chirality.  Further the coefficient of the n=2 Landau
wave function is a universal function of $t=T/T_{c}$.

\section{Vortex lattice}

A relatively large C (i.e. $C > \sqrt{\frac{3}{2}}-1 \simeq 0.22474$)
appears to be characteristic of triplet superconductivity with chirality
$\pm 1$, as observed in Sr$_{2}$RuO$_{4}$.  This suggests strongly that the
square vortex lattice is more stable than the usual hexagonal lattice except
in the immediate vicinity of the superconducting transitions at $T_{cA}$ and
$T_{cB}$.
\\
In order to examine the vortex lattice structure we calculate the Abrikosov
parameter \cite{9,10}
\bea
\beta_{A} &=& \frac{\left\langle|\Delta({\bf r})|^{4}\right\rangle}
{(\left\langle|\Delta({\bf r})|^{2}\right\rangle)^{2}}
\eea
For both the A and B phases in a magnetic field ${\bf H} \parallel {\bf c}$,
we obtain
\bea
\beta_{A} = \sqrt{\frac{R}{2\pi}}\sum_{n,m}(-1)^{nm}e^{-\frac{\pi}{2}R(n^{2}
+m^{2})} \times (1+ 2C)^{-2} \times \int_{-\infty}^{\infty} \, dx e^{-x^{2}}
(1+C^{2}(f(x+x_{1})+ f(x-x_{1})) \times \nonumber \\ 
\!\!\!\!\!\!\!\!\!\!\!\!\!\!\!\!\!\!\!\!\!(f(x+x_{2})+f(x-x_{2}))
+C^{4}f(x+x_{1})f(x-x_{1})f(x+x_{2})f(x-x_{2}))
\eea
where $f(x)=x^{2}-1$,\,$ R=\tan(\frac{\theta_{0}}{2}),$ 
$x_{1}=\sqrt{\frac{\pi R}{2}}(n-m)$, $x_{2}=\sqrt{\frac{\pi R}{2}}(n+m)$ and
$\theta_{0}$ is the apex 
angle of the vortex lattice.  Here $R=\sqrt{3}$ corresponds
to the regular hexagonal lattice, while $R=1$ is the square lattice.  The
above expression is the same as in the chiral p-wave superconductor discussed
in \cite{9}.  As is
well known, for the most stable lattice $\beta_{A}$ takes the minimum value.
From the analysis of the chiral p-wave superconductor \cite{9} we conclude
that the square vortex lattice is the most stable for $T < 0.95 T_{cA}$ and
$T < 0.95 T_{cB}$ in the A and B phase respectively.

Of course the present analysis applies only in the vicinity of $H=H_{c2}(T)$.
In order to find the phase boundary between the square vortex and the
hexagonal vortex lattice a parallel analysis, as was done for the vortex
lattice for d-wave superconductivity 
by Shiraishi et al \cite{15}, is required.  But we expect the 
transition line should be around $\kappa^{-1}H_{c2}(T)$ where $\kappa$ is
the Ginzburg-Landau parameter of PrOs$_{4}$Sb$_{12}$.  Making use of the
magnetic penetration depth determined by Chia et al \cite{chia} we can extract $\kappa
\simeq 30.0$.  This means that the square vortex lattice should be visible for
H= 0.1 T in an ideal crystal.  As already
mentioned Huxley et al \cite{16} studied the vortex lattice
via small angle neutron scattering (SANS), but saw only a distorted hexagonal
lattice.  In a magnetic field ${\bf H} \parallel [001]$, most of the nodal
excitations in both the A and B phases of PrOs$_{4}$Sb$_{12}$ are moving
parallel to the [100] and [010] directions.  Hence a square vortex lattice
aligned parallel to the crystal axis will minimize the quasi-particle
energy in the vortex state.

\section{Concluding Remarks}

We have analyzed the upper critical field H$_{c2}$ for the proposed p+h-wave 
superconductor PrOs$_{4}$Sb$_{12}$ \cite{6} and compared with the observed H$_{c2}(t)$
and H$^*(t)$.  The present model reproduces H$_{c2}$ without any adjustable
parameters.  Also the H$_{c2}$(t) observed in the single phase crystals
indicates the single phase is most likely the A phase.  Also H*(t)
appears to belong to the A phase.  Does this suggest inhomogeneity?  Also the present analysis predicts
that the square vortex lattice is favored when the magnetic field is
parallel to one of the crystal axes.  So far no strong correlation between
the superconducting transition temperature T$_{cA}$ and T$_{cB}$ and RRR \cite{measson}
has been observed.  This could be interpreted as meaning 
that disorder has little effect.  However,
from the early analysis of chiral p-wave superconductivity, we conclude that
$H_{c2}(T,\Gamma)/H_{c2}(0,\Gamma)$ is rather universal. 
An open question is whether a high-quality single crystal with $RRR > 100$
would show a single or double phase transition.  Clearly such a study is highly desirable.

{\bf Acknowledgements}

We thank Marie-Aude Measson for her generosity in sharing with us her
unpublished experimental data \cite{measson,measson2} and wealth of
new ideas.  Without her help this paper would not be completed in the
present form.  We have also benefitted from discussions
with Stephan Haas and Koichi Izawa.  
One of us (HW) acknowledges support from the Korean
Research Foundation through grant No. R05-2004-000-10814.

\end{document}